# SECRET SHARING FOR *n*-COLORABLE GRAPHS WITH APPLICATION TO PUBLIC KEY CRYPTOGRAPHY

Kamil Kulesza and Zbigniew Kotulski
Institute of Fundamental Technological Research, Polish Academy of Sciences
ul.Świętokrzyska 21, 00-049 Warsaw, Poland
E-mail: kkulesza@ippt.gov.pl; zkotulsk@ippt.gov.pl

**ABSTRACT**

*At the beginning some results from the field of graph theory are presented. Next we show how to share a secret that is proper* n-*coloring of the graph, with the known structure. The graph is described and converted to the form, where colors assigned to vertices form the number with entries from* $Z_n$. *A secret sharing scheme (SSS) for the graph coloring is proposed. The proposed method is applied to the public-key cryptosystem called "Polly Cracker". In this case the graph structure is a public key, while proper 3-colouring of the graph is a private key. We show how to share the private key. Sharing particular* n-*coloring (color-to-vertex assignment) for the known-structure graph is presented next.*

## 1. INTRODUCTION

Problems of proper vertex coloring for an arbitrary graph, with minimal set of colors are known to be NP [10]. Decisions problems about the graph coloring (e.g., on the graph chromatic number) are of NP class as well. Even simple problems like finding a 3-colouring for 3-colourable graph are known to be NP-hard. The latest is used in the graph based implementation of the public-key cryptosystem "Polly Cracker".

Public-key cryptography was pioneered by Diffie and Hellman [4]. Other important contributions came from (to name a few): Rivest, Shamir and Adleman [14] proposed RSA cryptosystem, ElGamal [5] build cryptosystem using the discrete logarithm problem, Koblitz [9] constructed public-key cryptosystem using elliptic curves. The public-key (asymmetric) cryptosystems use two different keys, opposed to private-key (symmetric) cryptosystems. In the private-key cryptosystems knowledge of one of the keys (nevermind, encryption or decryption) allows determination of the other from the pair, while in public-key cryptosystems knowing one of the keys, does not allow to determine the other. So, the keys are asymmetric, that allows to publish one key (public key) without compromising the other one (private key). Such algorithms provide much greater flexibility than traditional symmetric cryptosystems. They have two possible modes of operation:

a. *Secrecy*: message encrypted with the public key, can be decrypted only by the private key holder, hence message (plaintext) is kept secret and protected;
b. *Authenticity*: only private key holder can encrypt the message *m*, that can be read by anyone using the public key, hence identity of the private key holder is authenticated (protected).

Split control of the keys yields additional features for public-key cryptosystems. It can be implemented by means of secret sharing. Consider two instances with applications to both operation modes of the public-key cryptosystem:

a. Sharing the private key. The authorized entities must cooperate to recover the private key. *Secrecy:* message *m* encrypted using the public key can be decrypted only once the private key is recovered. *Authenticity:* recovered private key is used to encrypt the message. Hence, with the help of the public key it is possible to carry out authentication procedure and prove that message was encrypted by the authorized entities.
b. Sharing the public key. Various modifications of secret sharing schemes can be used (see [11]). To illustrate the point consider pre-positioned secret sharing schemes. In [11] such schemes were defined as follows : „All necessary secret information is put in place excepting a single (constant) share which must later be communicated, e.g. by broadcast, to activate the scheme". *Secrecy:* message *m* can be encrypted only when the public key is known. *Authenticity:* again message *m* encrypted using the private key can be authenticated only when the public key is recovered.

Thus, it is possible to design structures with various level of openness and privacy. In the broader perspective, split control of the keys in the public-key cryptosystems can be seen as the new paradigm that sets intermediate states between two opposite realms of public and private. Such structures can find applications not only, where the public-key cryptosystems and secret sharing are applied nowadays. Combined, can enter fields like: managing complex processes on the financial markets or multiparty decisions in the corporate governance field.

The secret sharing allows splitting a secret into different pieces, called shares, which are given to the participants. Only certain group (the authorized set of participants) can recover the secret. Secret sharing schemes were independently invented by George Blakley [2] and Adi Shamir [15]. Many schemes were presented since, for instance: Asmuth and Bloom [1], Brickell [3], Karin-Greene-Hellman (KGH method) [7].

Since the last one will be used in the examples through the paper, its description is provided below.

KGH is a simple and elegant method with striking similarity to the Vernam cipher, see [7]. The secret is a vector of $\eta$ numbers $S_\eta = \{s_1, s_2, ..., s_\eta\}$. Some modulus *k* is chosen, such that $k > \max(s_1, s_2, ..., s_\eta)$. All *t* participants are given shares that are $\eta$-dimensional vectors $S_\eta^{(j)}, j = 1, 2, ..., t$ with elements in $Z_k$. To retrieve the secret they have to add the vectors component-wise in $Z_k$.

For $k = 2$, KGH method works like $\oplus$ (xor) on $\eta$-bits numbers, much in the same way like Vernam one-time pad. If *t* participants are needed to recover the secret, adding *t*-1 (or less) shares reveals no information about secret itself. Interesting feature of KGH is that when certain vectors $S_\eta^*$ are excluded (not allowed) from the set of possible secret values, method remains equally secure. Again, having *t*-1 parts (or less) of the secret reveals no information about the secret itself. KGH with excluded vectors is referred as KGHe. Certainly, for same $\eta$ (vector length) the cardinality of the "secret space" is smaller for KGHe than for KGH.

In practice, it is often needed that only certain specified subsets of the participants should be able to recover the secret. The authorized subset is a subset of secret participants that are able to recover secret. The access structure describes all the authorized subsets. To design the access structure with required capabilities, the cumulative array construction can be used, for details see, for example, [6]. Combining cumulative arrays with KGH method, one obtains an implementation of the general secret sharing scheme (see, e.g., [12]). While designing such an implementation, one can introduce required capabilities not only in terms of the access structure but also others, like security (e.g., perfectness), see [11], [17].

The outline of the paper is the following: Section 2 briefly summarizes results from graph theory needed further in the paper. First procedure to convert any graph into the form convenient for the secret sharing (see section 2.1) is given, then graph *n*-coloring results needed further in the text are presented. In section 3 we describe graph coloring based implementation of "Polly Cracker" public-key cryptosystem. Method to share private-key in the described "Polly Cracker" implementation is introduced in the next section. Procedure for sharing particular graph coloring in the graph with known structure follows (section 5).

## 2. GRAPHS COLORING RESULTS

Notation:
$G(V,E)$ is the graph, where *V* is set of vertices and *E* is set of edges, with *e* edges and *v* vertices,

$v_i$ denotes *i*th vertex of the graph, $v_i \in V$,

$K_n$ is the complete graph on *n* vertices (the graph which has edges connecting all vertices),

$\chi(G)$ is the chromatic number of graph *G*

(the minimal number of colors needed for vertex coloring of the graph). In this section graphs $G$ with $\chi(G) = n$ are considered, unless stated otherwise. All the examples are given for 3-colorable graphs.

### 2.1 Graph description

Graph $G$ is described by the square adjacency matrix $\mathbf{A} = [a_{ij}], i, j = 1, 2, ..., m$. The elements of $\mathbf{A}$ satisfy:
- for $i \neq j$, $a_{ij} = 1$ if $v_i v_j \in E$ (vertices $v_i$, $v_j$ are connected by an edge) and $a_{ij} = 0$, otherwise;
- for $i = j$, $a_{ii} = n$, where $n \in Z_k$ is the number of color assigned to $v_i$. In $Z_k$, $k \geq \chi(G)$ denotes the number of colors that can be used to color vertices of $G$ (in other words, $k$ is the size of the color palette).

In the case that the graph coloring is not considered, $k=1$, and all entries on $\mathbf{A}$'s main diagonal are zero.

*Example 1*

Take the graph $G$ with 4 vertices, colored with 3 colors:

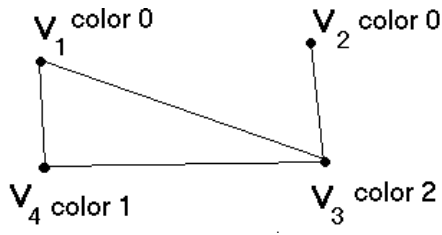

The adjacency matrix of the graph $G$ (only the graph structure, no colors) is:

$$\begin{array}{c|cccc} & v_1 & v_2 & v_3 & v_4 \\ \hline v_1 & 0 & 0 & 1 & 1 \\ v_2 & 0 & 0 & 1 & 0 \\ v_3 & 1 & 1 & 0 & 1 \\ v_4 & 1 & 0 & 1 & 0 \end{array}$$

while the whole adjacency matrix $\mathbf{A}$ with encoded coloring is

$$\begin{array}{c|cccc} & v_1 & v_2 & v_3 & v_4 \\ \hline v_1 & 0 & 0 & 1 & 1 \\ v_2 & 0 & 0 & 1 & 0 \\ v_3 & 1 & 1 & 2 & 1 \\ v_4 & 1 & 0 & 1 & 1 \end{array}$$
∎

Coloring and the chromatic number are integral properties of any graph. Given the graph, it is always possible to find its $n$-coloring and chromatic number. In the general case, this is the problem of NP class, see [10]. Both properties cannot be separated from the graph structure itself. Hence, one may try to use them for own advantage.

It is worthy to note that color encoding provided allows increasing number of graph $G$ colorings. Take $n = \chi(G) < k$. Then for every proper vertex coloring of the graph $G$ there are $\binom{k}{n}$ colorings of the graph $G$ with $n$ colors from $k$-color palette.

### 2.2 Coding the matrix A

$\mathbf{A}$ is a symmetric matrix, hence having all the entries on the main diagonal and all the entries below main diagonal, one can describe whole matrix (and as the result graph $G$). Thus it can be written as the sequence $a_{21}a_{31}a_{32}a_{41}a_{42}a_{43} ... a_{m(m-1)}a_{11} a_{22} ... a_{mm}$, where the first binary part ($a_{21}a_{31}a_{32}a_{41}a_{42}a_{43} .. a_{m(m-1)}$) corresponds to all the entries below main diagonal, while second decimal one ($a_{11} a_{22} ... a_{mm}$) to the main diagonal itself.

*Example 1* (continuation)
Coding matrix $\mathbf{A}$ we obtain

| $a_{21}$ | $a_{31}$ | $a_{32}$ | $a_{41}$ | $a_{42}$ | $a_{43}$ | $a_{11}$ | $a_{22}$ | $a_{33}$ | $a_{44}$ |
|---|---|---|---|---|---|---|---|---|---|
| 0 | 1 | 1 | 1 | 0 | 1 | 0 | 0 | 2 | 1 |

that yields $m = 0111010021$ ∎

### 2.3 Vertex types in graph $n$-coloring

In general graph $G$ $n$-coloring is equivalent to partitioning it into $n$ sets of vertices, such that vertices in one set are not connected (hence $n$-coloring of such a graph), see [10].

**Definition**. The degree of freedom of the vertex in graph $G$ for particular coloring is numbers of colors that can be assigned to that vertex in graph $n$-coloring. Alternatively one can compare all colors excluded for particular vertex (vertex is connected with vertices having such colors assigned) with all colors available for the coloring.

In the graph G ($\chi(G) = n$) every vertex in the graph can be assigned one of the following types:

*Type I*: fixed vertex with degree of freedom equals 1. In all possible graph colorings only one color remains for such a vertex. For instance, check any vertex in *n*-coloring of $K_n$ graph.

*Type II*: fixed vertex with degree of freedom equal to *y* ($y < n$; $y \in N$). In all possible graph colorings, *y* colors remain available for such a vertex.

**Example 2** of vertex with the degree of freedom *y*=2 and *n*=3. On the drawing numbers next to vertices denote assigned coloring, while corresponding matrix follows.

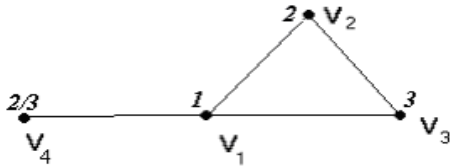

|       | $v_1$ | $v_2$ | $v_3$ | $v_4$ |
|-------|-------|-------|-------|-------|
| $v_1$ | 1 | 1 | 1 | 1 |
| $v_2$ | 1 | 2 | 1 | 0 |
| $v_3$ | 1 | 1 | 3 | 0 |
| $v_4$ | 1 | 0 | 0 | 2/3 |

*Type III*: slack vertex with the variable degree of freedom. The degree of freedom depends on the particular graph coloring. ∎

**Example 3** of slack vertex ($v_5$) with variable degree of freedom for *n*=3. On the drawing numbers next to vertices denote assigned coloring, while corresponding matrices follow.

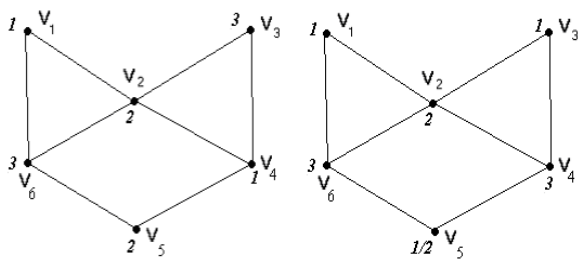

|       | $v_1$ | $v_2$ | $v_3$ | $v_4$ | $v_5$ | $v_6$ |
|-------|-------|-------|-------|-------|-------|-------|
| $v_1$ | 1 | 1 | 0 | 0 | 0 | 1 |
| $v_2$ | 1 | 2 | 1 | 1 | 0 | 1 |
| $v_3$ | 0 | 1 | 3 | 1 | 0 | 0 |
| $v_4$ | 0 | 1 | 1 | 1 | 1 | 0 |
| $v_5$ | 0 | 0 | 0 | 1 | 2 | 1 |
| $v_6$ | 1 | 1 | 0 | 0 | 1 | 3 |

|       | $v_1$ | $v_2$ | $v_3$ | $v_4$ | $v_5$ | $v_6$ |
|-------|-------|-------|-------|-------|-------|-------|
| $v_1$ | 1 | 1 | 0 | 0 | 0 | 1 |
| $v_2$ | 1 | 2 | 1 | 1 | 0 | 1 |
| $v_3$ | 0 | 1 | 1 | 1 | 0 | 0 |
| $v_4$ | 0 | 1 | 1 | 3 | 1 | 0 |
| $v_5$ | 0 | 0 | 0 | 1 | 1/2 | 1 |
| $v_6$ | 1 | 1 | 0 | 0 | 1 | 3 |

∎

### 2.4 Reducibility

Type I vertices form disjoint subgraphs in graph *G*. For every such a subgraph, minimal set of vertices that uniquely determine *n*-coloring (of the subgraph) can be found. Hence, these vertices can be reduced (in the coloring sense) to the smaller set.

The reduced structure is a minimal set of type I vertices that uniquely determine *n*-coloring for connected graph made of type I vertices.

**Example 4**

On the drawing numbers next to vertices denote assigned coloring, while corresponding matrix follows.

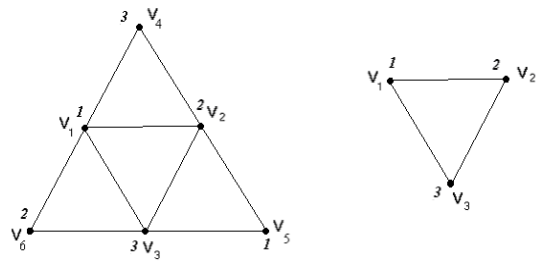

|       | $v_1$ | $v_2$ | $v_3$ | $v_4$ | $v_5$ | $v_6$ |
|-------|-------|-------|-------|-------|-------|-------|
| $v_1$ | 1 | 1 | 1 | 1 | 0 | 1 |
| $v_2$ | 1 | 2 | 1 | 1 | 1 | 0 |
| $v_3$ | 1 | 1 | 3 | 0 | 1 | 1 |
| $v_4$ | 1 | 1 | 0 | 3 | 0 | 0 |
| $v_5$ | 0 | 1 | 1 | 0 | 1 | 0 |
| $v_6$ | 1 | 0 | 1 | 0 | 0 | 2 |

reduces to (vertices $v_1 v_2 v_3$):

|       | $v_1$ | $v_2$ | $v_3$ |
|-------|-------|-------|-------|
| $v_1$ | 1 | 1 | 1 |
| $v_2$ | 1 | 2 | 1 |
| $v_3$ | 1 | 1 | 3 |

∎

### 2.5 Remarks on constructing required *n*-colorable graphs

Rigorous and formal treatment of the

subject would by far exceed the space limitations. Instead, only some ideas and sketches of algorithms will be presented.

**Using colors.** Knowing graph $G$, with $\chi(G) = n$, one is not restricted to using only $n$ colors for the given graph coloring. Graph $G$ can be properly colored with any $z$ colors ($n \leq z \leq |V|$, $z \in N$). Colors needed for graph coloring (for $\chi(G) = n$) can be chosen from much greater palette, e.g., $k$ possible colors. Then, the number of colors' combinations is simply $\binom{k}{n}$.

**Toolbox for checking chromatic number and constructing required graphs.** Some theoretical results can help in constructing proper graphs for the secret sharing purpose. Here we present some of them.

**Theorem.** *Every graph $G$ of $K_n$ configuration has $\chi(G) = n$.*

**Lemma.** *Every graph G having a subgraph of $K_n$ configuration has $\chi(G) \geq n$.*

These results set lower bound on $\chi(G)$ when constructing graph $G$.

**The Brook's theorem** (1941). *If the graph G is not an odd circuit or a complete graph, then $\chi(G) \leq d$, where d is the maximum degree of a vertex of G.*

The Brook's theorem sets the upper bound on $\chi(G)$ when constructing graph $G$. It is useful when building graph $G$ from smaller blocks.

**Theorem.** *When two disjoint graphs $G_1$ ($\chi(G_1) = n_1$) and $G_2$ ($\chi(G_2) = n_2$) are linked by any number of edges to form graph G, the following holds:*

$$\max(\chi(G_1), \chi(G_2)) \leq \chi(G) \leq \chi(G_1) + \chi(G_2).$$

One can also use the idea of vertex types (see section 2.3) to build an adequate graph. Consider two examples of such structures:
- When graph is built from type I and II vertices, result is straight forward (although resulting graphs can have higher chromatic number, then one used to define vertex types).
- Starting from the graph that vertex types are determined and step-by-step adding vertices and edges in the way that vertex type assessment remains feasible (when preceding graph type assignment is known).

## 3. POLLY CRACKER

Consider the particular implementation of public-key cryptosystem "Polly Cracker", that uses graph 3-coloring (see [8]). Although successful attacks on the general Polly Cracker was described in [16], we decided to use it as general illustrative example. We find it as convenient vehicle for presenting more general concepts and opportunities resulting from sharing of graph properties.

The general idea behind graph based implementation of the Polly Cracker scheme is as follows:

a. Construct polynomials over finite field $F$.

b. Take an arbitrary vector $z \in F^n$ as a private key and the subset $B = \{q_i\}$ of the polynomials over finite field $F$, such that, for every $i$, $q_i(z) = 0$, as a public-key.

c. Encrypt a message $m$ obtaining cipher polynomial $C$ using the public-key (a randomly chosen element generated by $B$).

d. Message $m$ can be decrypted by finding the value of polynomial $C$ at $z$.

Having described public-key cryptosystem "Polly Cracker", one can move to its special case based on graph 3-coloring. The problem of graph 3-coloring is NP class (see [10]). To formulate the cryptosystem in terms of the graph theory, we introduce as the public-key the graph $G(V,E)$, that is the graph with the set of vertices $V$ and the set of edges $E$ and, as the private key, the proper 3-coloring of the graph using colors $s \in \{0,1,2\}$ and the map assigning $v_i \mapsto s$ for $v_i \in V$, according to graph 3-coloring rule.

Once graph 3-coloring is known, the base $B = B(G)$ is constructed. $B$ is constructed from a polynomial derived from the variables $\{t_{v,s}\}$, and $B = B_1 \cup B_2 \cup B_3$ for
$B_1 = \{t_{v,0} + t_{v,1} + t_{v,2} - 1 : v \in V\}$
$B_2 = \{t_{v,s} t_{v,p} : v \in V, s \leq p \in \{0,1,2\}\}$,
$B_3 = \{t_{v,s} t_{w,s} : e(v,w) \in E\}$

Then, the zero point of all polynomials from $B$ can be computed by taking $t_{v,s} = 1$, if the vertex $v$ has color $s$, and 0 otherwise.

In a similar way other graph based "Polly Cracker" schemes can be constructed. One of the examples can be "perfect code"

graph described in [8].

All these implementations, like described above graph 3-coloring system, have the following features:
a. Knowing $G(V,E)$ is equivalent to knowing subset $B = \{q_i\}$ of polynomials over finite field $F$.
b. Knowing the NP-class problem (resulting from the graph structure) is equivalent to knowing vector $z \in F^n$.
c. Encryption takes place like in general "Polly Cracker" scheme.
d. To decrypt message $m$, a value of the received polynomial (derived from the graph $G(E,V)$ structure) at $z$ should be calculated.

The private key is a proper 3-coloring of the graph. It is important to note that any proper 3-coloring of the graph can be the private key. Hence, for the given graph $G$ (the public key) there are usually more then one (in fact, usually much more) private keys.

## 4. SHARING POLLY CRACKER'S PRIVATE KEY

As described in the section 2.2, vertex coloring of the graph $G$ can be written as the sequence $a_{11}\ a_{22}\ ...a_{mm}$ with entries from the main diagonal of matrix **A**. To share graph's coloring is to share this sequence (vector). For this purpose all secret sharing methods suitable to share number can be applied.

However one should note that information contained in the graph structure severely limits the secret space. The particular secret space needs to be individually examined.

It should be emphasized that in general case one can share only partitioning graph's vertices into $n$ sets (proper $n$-coloring for the graph), where $n = \chi(G)$, not a particular color-to-vertex assignment. It is due to the fact that in linear secret sharing scheme any secret participant can modify her share adding component-wise in $Z_k$ a constant to every digit in the number. In such a case:
a. Particular color-to-vertex assignment will be modified.
b. Partitioning graph's vertices into $n$ sets (proper $n$-coloring for the graph) will remain valid.

The algorithm for the case when coloring with particular color-to-vertex assignment is securely shared is described in the section 5. The other possibility would arise from using nonlinear secret sharing schemes, see [13].

To share the Polly Cracker's private key (for the implementation presented in section 3) one needs to share proper 3-colouring for the graph, which can be done by any of existing secret sharing methods (see [11], [12]). To illustrate this process authors use KGH. It is to be shown that KGHe (all invalid 3-colorings are excluded) method can be used for this purpose. Finding proper 3-coloring for the graph is the problem of NP class, see [10], hence finding the private key for Polly Cracker is a difficult task.

Now, assume that KGHe requires co-operation of $t$ participants to recover the secret. Let $t$-1 participants to pool their shares. Then, the result they receive can be changed into any element from the secret space (a vector of $\eta$ numbers), when the lacking share of the secret is xored ($\oplus$). Set of possible values of the last (unknown) share can be restricted only when proper 3-colorings of the graph are know. But this is the secret that is being shared !!!

When the secret is not known, xor ($\oplus$) of $t$-1 shares reveals no information about the secret number and does not help to find the private key.

It is interesting to note that, in order to recover the private key, secret shares have to add up ($\oplus$) to any proper graph 3-coloring. While designing the scheme's implementation (graph for the public key) one can usually compute its proper colorings much easier (see section 2.5) than from average ready-made graph. This fact opens new possibility for designing access structures. For instance, although different sets of authorized participants retrieve different secrets, so secret sharing scheme looks like multi-secret threshold schemes, see [11], each of different secrets has the same functionality, being a valid private key for a particular Polly Cracker implementation.

## 5. SHARING PARTICULAR GRAPH COLORING IN THE GRAPH WITH

**KNOWN STRUCTURE**

The graph $G$, build of type I and type II vertices (see section 2.3), will be used to propose a secret sharing method. Type assignment is equivalent to partitioning graph $G$ into $n$ sets of vertices, such that vertices in one set are not connected (hence, finding $n$-coloring of such a graph). Particular graph $n$-coloring (color-to-vertex assignment), with colors taken from predefined $k$-color palette, is the secret to be shared.

There are two separate pieces of the secret that can be shared independently: type I vertices coloring and type II vertices coloring. If only one of the parts is reconstructed, the rest still remains a secret. Certainly, it is possible to find a finite set of possible secret values, but its cardinality (size) can be decided during the implementation, to meet required security level.

First, method to share coloring for both vertex types will be supplied. It is strongly recommended to read this part of the paper simultaneously with example 5 that follows. References to the particular steps of the example are given. Although proposed approach works with many secret sharing methods, again KGH is used in discussion.

**5.1 Sharing the coloring of type I vertices**

To start the procedure, find reduced structure for type I vertices in the graph $G$ (see section 2.4). Each vertex $v_i$ from the reduced structure is assigned a color $s$ from $Z_k$, $k \geq \chi(G) = n$ (see section 2.4). Certainly, only $n$ out of $k$ colors can be used at once. The reduced structure can be written as the vector of $r$ numbers $S_r = \{s_1, s_2, ..., s_r\}$ (where $s_j \in Z_k, j = 1, 2, ..., r$, are colors assigned to the vertices $v_i$ in the reduced structure; vertices are written in ascending order with respect to the index $i$).

First case, when $\chi(G) = n = k$ will be discussed. In this situation vertices in the reduced structure are assigned colors from $Z_n$. Let's name such an assignment "$Z_n$ encoding for the graph $G$". There are at least $n!$ of $Z_n$ encodings for the graph $G$. Clearly, an attacker can easily determine $n$-coloring of vertices from reduced structure, but will not know particular $Z_n$ encoding for the graph $G$. To share it KGHe can be used. "Full" KGH cannot be applied, because vertices in the reduced structure, which are linked by the common edge, must have different colors. Using the same reasoning as in the section 4, it can be shown that xor ($\oplus$) of any unauthorized set of shares (even just below the threshold) reveals no information about secret and does not help to find $Z_n$ encoding for the graph $G$.

Second case arise when $k > \chi(G) = n$. In such a situation, secret that is being shared, consists of particular $\binom{k}{n}$ colors combination and their particular permutation (color-to-vertex assignment for every $v_i$). In such a case the following routine is applied:

**Algorithm 1:** *for coloring vertices of type I*

1. Proper $n$-coloring of the graph $G$ is known.
2. The reduced structure for type I vertices in the graph $G$ (see section 2.4) is found.
3. Let's name all numbers in the particular $\binom{k}{n}$ colors combination, used to color graph $G$, as "particular $n$ colors from $Z_k$". First put them in the ascending order applying ordering principle for $N$ (natural numbers). Next particular $n$ colors from $Z_k$ are assigned numbers from the set $\{0, 1, 2, ..., n-1\}$. This is done by staring from the smallest element in the set of the particular $\binom{k}{n}$ colors combination and using consecutive numbers from set $\{0, 1, 2, ..., n-1\}$ to enumerate consecutive colors (numbers). Once this is done, the mapping between particular $n$ colors from $Z_k$ and $Z_n$ is established.
4. Once the mapping is known, the $Z_n$ encoding for the graph $G$ is determined.
end // *for coloring vertices of type I*

**Discussion:** When the $Z_n$ encoding for the graph $G$ is known and can be shared using KHGe as described above. To see this routine at work, check step 1 in the example 5.

**5.2 Sharing the coloring of type II vertices**

Technical remark: When $k > \chi(G) = n$, if type II vertices coloring (one piece of the secret) was found, particular $\binom{k}{n}$ colors combination encoded in type I vertices would not be a secret any more.

To avoid it, colors from $Z_k$ should be replaced by the colors from $Z_n$. This is done using $Z_k$ to $Z_n$ mapping found for type I vertices above. So, the case when $k > \chi(G) = n$, can be reduced to the case $\chi(G) = k = n$.

Having type II vertices assigned colors from $Z_n$, arise some other problems :

1. When type II vertices coloring (one piece of the secret ) is known, one can deduce type I vertices coloring (or, at least, severely limit the number of available possibilities).

2. No good routine is known for quantitative analysis of the method's security parameters.

To address these problems, type II vertices have to be converted into more convenient form.

For type II vertices define:

a. $n_i$ is the number of colors excluded for the particular vertex $v_i$. It is obtained by checking vertices of type I that are linked to $v_i$. Clearly, every color that is assigned to any vertex linked to $v_i$ is excluded from the list of available colors.

b. $l_i$ is the number of the colors available for the vertex $v_i$, $l_i = n - n_i$.

c. $Z_{l_i} = \{0,1,2,...,l_i - 1\}$.

d. $C_i$ is the set of $l_i$ colors from $Z_n$, that are available for the vertex $v_i$.

e. $w$ is the number of type II vertices in the graph $G$.

For practical instances of the terms defined above see step 2 in the example 5.

Due to the fact that graph $G$ is properly $n$-colored using colors from $Z_n$, each vertex $v_i$ has assigned a color from $C_i$. Note that $|C_i| = |Z_{l_i}|$, hence one-to-one mapping between $Z_{l_i}$ and $C_i$ can be defined.

**Algorithm 2:** *for $Z_{l_i}$ and $C_i$ one-to-one mapping*

For every particular type II vertex $v_i$ do:
a. Put elements of $Z_{l_i}$ and $C_i$ in ascending order applying ordering principle for $N$.
b. Once elements in $C_i$ and $Z_{l_i}$ are ordered, they can be labeled (enumerated). This is done by starting from the smallest element in $C_i$ and using consecutive numbers from $\{0,1,2,...,l_i - 1\}$ for consecutive ordered elements from $C_i$.

c. When described in previous point (b) routine is completed, mapping between $C_i$ and $\{0,1,2,...,l_i - 1\}$ is found. Hence, mapping between $C_i$ and $Z_{l_i}$ is also known. This allows to express $v_i$ coloring in terms of colors from $Z_{l_i}$, for each $v_i$.

end // *for $Z_{l_i}$ and $C_i$ one-to-one mapping*

**Discussion:** To see algorithm at work, consult step 3 in the example 5.

Note that a color from $Z_n$ assigned to particular vertex $v_i$ of type II, can be encoded by any number from $Z_{l_i}$ with equal probability. So, type II vertices coloring using $Z_{l_i}$ does not provided any information on $Z_n$ encoding for the graph $G$. Hence, it is obvious that such a single number $s_i$ (particular $v_i$ color taken from $Z_{l_i}$), can be shared using KGH. The sequence of $w$ such numbers, that yields the vector $S_w$ can be also shared by KGH. This concludes the part concerning sharing the coloring of type II vertices.

**5.3 Interaction between secrets resulting from different types of vertices**

Now is time for few comments on situations, when one of the pieces of the secret (coloring for one of vertex types) is known. To describe it, two cases will be discussed:

1. Secret information for type II vertices was recovered. Then, there are at least $\binom{k}{n}n!$ particular color assignments for type I vertices. No information about particular $\binom{k}{n}$ combination or $Z_n$ encoding for the graph $G$ can be derived from known coloring of type II vertices.

2. Secret information for type I vertices was recovered. Then, there are $\prod_{i=1}^{\eta} a_i$ possibilities to assign available colors to type II vertices, where $a_i = \begin{cases} l_i & \text{for type II vertices} \\ 1 & \text{otherwise} \end{cases}$. Knowing type I vertices coloring one can only determine available colors for each of type II vertices, but has no information about which of the colors is chosen.

***Example 5*** of sharing particular graph coloring in the graph with known structure.

Take the 3-colorable graph, colors are assigned from $Z_{10}$.

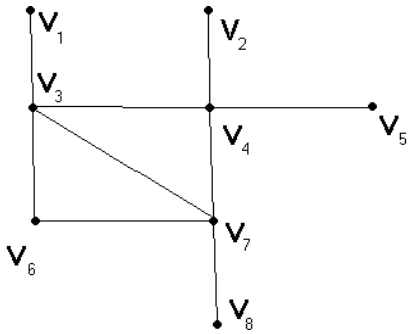

The matrix **A** for the given graph is:

|       | $v_1$ | $v_2$ | $v_3$ | $v_4$ | $v_5$ | $v_6$ | $v_7$ | $v_8$ |
|-------|-------|-------|-------|-------|-------|-------|-------|-------|
| $v_1$ | 5 | 0 | 1 | 0 | 0 | 0 | 0 | 0 |
| $v_2$ | 0 | 2 | 0 | 1 | 0 | 0 | 0 | 0 |
| $v_3$ | 1 | 0 | 2 | 1 | 0 | 1 | 1 | 0 |
| $v_4$ | 0 | 1 | 1 | 5 | 1 | 0 | 1 | 0 |
| $v_5$ | 0 | 0 | 0 | 1 | 6 | 0 | 0 | 0 |
| $v_6$ | 0 | 0 | 1 | 0 | 0 | 5 | 1 | 0 |
| $v_7$ | 0 | 0 | 1 | 1 | 0 | 1 | 6 | 1 |
| $v_8$ | 0 | 0 | 0 | 0 | 0 | 0 | 1 | 5 |

and the matrix **A** coding yields:
01001100010010000110100000152256565

*step 1:*

Type I vertices: $v_3 v_4 v_6 v_7$ yield the reduced structure: $v_3 v_4 v_7$. Mappings $Z_{10}$ into $Z_3$ are defined: $v_3=0$, $v_4=1$, $v_7=2$ (all in $Z_3$), yield $2\to 0$, $5 \to 1$, $6 \to 2$. Hence, $Z_3$ encoding for the graph is $v_3 v_4 v_7 \to 012$.

*step 2:*

Type II vertices: $v_1$ $v_2$ $v_5$ $v_8$, colors assignment from $Z_{10}$ to $Z_3$ is: $v_1=5 \to 1$, $v_2=2 \to 0$, $v_5=6 \to 2$ $v_8=5 \to 1$. So, $v_1 v_2 v_5 v_8$ corresponds to 1021 in $Z_3$ encoding.

Sets of numbers from $Z_3$ excluded for particular vertex: $v_1$ {0}, $v_2$ {1}, $v_5$ {1}, $v_8$ {2}.

Sets of numbers from $Z_3$ available for the particular vertex: $v_1 \in C_1=\{1,2\}$, $v_2 \in C_2=\{0,2\}$, $v_5 \in C_5=\{0,2\}$, $v_8 \in C_8=\{0,1\}$.

In this example $l_i=2$ for each of vertices: $v_1$ $v_2 v_5 v_8$, hence $Z_{l_i} = Z_2$.

*step 3:*

For $v_1$ mapping $Z_3 \to Z_2$ is defined {1,2} $\to$ {0,1}, hence $1\to 0$ and $2\to 1$.

For $v_2, v_5$ mapping $Z_3 \to Z_2$ is defined {0,2} $\to$ {0,1}, hence $0\to 0$ and $2 \to 1$.

For $v_8$ mapping $Z_3 \to Z_2$ is defined {0,1} $\to$ {0,1}, hence $0\to 0$ and $1 \to 1$.

Finally, $v_1 v_2 v_5 v_8$ corresponds to 0011 in $Z_2$ encoding.

Hence, for the given graph $G$ the number to be shared is 0011256, where 0011 corresponds to type II vertices and 256 corresponds to the reduced structure of type I vertices. The first part of the number can be shared by "full" KGH. The second part of the number has to be shared by KGHe, since some of the vertex pairs in the reduced structure must have different colors.

Now it is time to calculate numbers for described above cases when one of the pieces of the secret is known.

Case 1. $\binom{10}{3} 3! = 720 = 6!$

Case 2. There are 4 type II vertices: $v_1 v_2 v_5$ $v_8$ and for each $l_i=2$, hence $\prod_{i=1}^{8} a_i = 2^4$

In both cases numbers do not seem to be impressive, but in a general case they can be made as big as required during the implementation. ∎

To recover secret the following algorithm has to be applied. It works independently whether $k > \chi(G)$ or $k=\chi(G)=n$. In the later case just substitute $k$ by $n$ in the routine description and skip all references to the mapping $Z_k$ and $Z_n$.

**Algorithm 3**: *secret recovery:*

1. Authorized participants pool the shares → secret number for the graph (0011256 in example above) is obtained.
2. Secret number is used to:
a. establish colors from $Z_k$ for the reduced structure, this also yields coloring using $Z_n$ colors (remember that both subsets are ordered in *N*),
b. once the reduced structure coloring is

known, coloring for all type I vertices is established,

c. establish colors from $Z_{l_i}$ for vertices of type II.

3. Using coloring of type I vertices, $C_i$ for every type II vertex $v_i$ is found.
4. For each vertex of type II determine its $Z_n$ color, using $Z_{l_i}$ and $C_i$ (remember that both subsets are ordered in *N*).
5. For all type II vertices, $Z_k$ colors are assigned using their $Z_n$ colors and known $Z_k$ to $Z_n$ mapping.

Upon completing procedure, the secret (particular *n*-coloring for the known graph, using *k*-color palette) is recovered.

end. // *secret recovery*

Remark I:

The method described in this section works in graphs with type I and II vertices. Addressing issue of vertices of type III, although easy in some special instances, seems to be difficult in general case and requires further research.

Remark II:

Restriction of the method to vertices of type I and II does not seem so harmful having in mind tools that are available for the designer of the graph and secret sharing scheme (see section 2.5)

## 6. CONCLUDING REMARKS AND FURTHER RESEARCH

In this paper, we have shown how to share graph vertex coloring. Although we used KGH scheme as the example, all secret traditional sharing methods, that are used to share secrets consisting of numbers, can be applied.

Further research will concentrate on sharing of other graph related properties (e.g., Hamiltonian paths, graphs isomorphism) for graphs with known structure.

## BIBLIOGRAPHY

[1] C. Asmuth and J. Bloom, A modular approach to key safeguarding, IEEE Trans. on Information Theory, IT-29 No.2(1983), pp.208-211.

[2] G.R. Blakley, Safeguarding cryptographic keys, In: Proc. AFIPS 1979 National Computer Conference, pp. 313-317. AFIPS, 1979.

[3] E.F. Brickell, Some ideal secret sharing schemes, Journal of Combinatorial Mathematics and Comb. Computing, 6(1989), pp. 105-113.

[4] W. Diffie and M. E. Hellman, New directions in cryptography, IEEE Trans. Inform. Theory, IT-22 (1976), pp. 644-654.

[5] T. ElGamal, A public key cryptosystem and a signature scheme based on discrete logarithms, IEEE Transactions on Information Theory, IT-31 (1985), pp. 469-472.

[6] M. Ito, A. Saito, and T. Nishizeki, Secret sharing scheme realizing general access structure, In: Proceedings IEEE Globecom '87, pp. 99-102. IEEE, 1987.

[7] E.D. Karnin, J.W. Greene, and M.E. Hellman, On secret sharing systems, IEEE Transactions on Information Theory, IT-29 (1983), pp. 35-41.

[8] N.Koblitz, Algebraic Aspects of Cryptography, Springer-Verlag, Berlin 1998.

[9] N. Koblitz. Elliptic curve cryptosytems, Math. of Computation, 48(177) (1987), pp .203-209.

[10] B. Korte, J.Vygen, Combinatorial Optimization, theory and algorithms, Springer-Verlag, Berlin 2000.

[11] A.J. Menezes, P. van Oorschot and S.C. Vanstone Handbook of Applied Cryptography, CRC Press, Boca Raton 1997.

[12] ] J. Pieprzyk ,T. Hardjono and J. Seberry Fundamentals of Computer Security. Springer-Verlag, Berlin. 2003.

[13] Pieprzyk J., Xian-Mo Zhang 2001. 'Nonlinear secret sharing immune against cheating'. *The 2001 International Workshop on Cryptology and Network Security,* pp. 154-161

[14] Ronald L. Rivest, Adi Shamir, and Leonard M. Adleman, A method for obtaining digital signatures and public-key cryptosystems. Comm. of the ACM, 21(1978), pp.120-126.

[15] A. Shamir, How to share a secret, Communication of the ACM, 22(1979),pp.612-613

[16] R. Steinwandt, W. Geiselmann, R. Endsuleit, Attacking a polynomial-based cryptosystem: Polly Cracker, International Journal of Information Security, 1 (2002), pp. 143-148.

[17] D.R. Stinson, "Cryptography, Theory and Practice", CRC Press, Boca Raton 1995.